\documentclass{article}

\usepackage{arxiv}

\usepackage[utf8]{inputenc} 
\usepackage[T1]{fontenc}    
\usepackage{hyperref}       
\usepackage{url}            
\usepackage{booktabs}       
\usepackage{amsfonts}       
\usepackage{nicefrac}       
\usepackage{microtype}      
\usepackage{lipsum}		
\usepackage{graphicx}
\usepackage{natbib}
\usepackage{doi}

\title{Software-based Security Framework for Edge and Mobile IoT}

\author{ \href{https://orcid.org/0000-0002-5351-5580}{\includegraphics[scale=0.06]{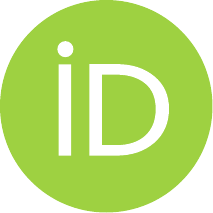}\hspace{1mm}José Cecílio} \\
	LASIGE, Departamento de Informática, \\
Faculdade de Ciências da Universidade Lisboa,\\
Campo Grande 016, 1749-016 Lisboa\\
	\texttt{jcecilio@ciencias.ulisboa.pt} \\
	\And
	\href{https://orcid.org/0000-0001-6311-9672}{\includegraphics[scale=0.06]{orcid.pdf}\hspace{1mm}Alan Oliveira de Sá} \\
	LASIGE, Departamento de Informática, \\
Faculdade de Ciências da Universidade Lisboa,\\
Campo Grande 016, 1749-016 Lisboa\\
	\texttt{aodsa@ciencias.ulisboa.pt} \\
	\AND
 \href{https://orcid.org/0000-0001-8792-959X}{\includegraphics[scale=0.06]{orcid.pdf}\hspace{1mm}André Souto} \\
	LASIGE, Departamento de Informática, \\
Faculdade de Ciências da Universidade Lisboa,\\
Campo Grande 016, 1749-016 Lisboa\\
	\texttt{ansouto@ciencias.ulisboa.pt} \\
}

\renewcommand{\headeright}{Preprint AEIC-2024}
\renewcommand{\undertitle}{Preprint AEIC-2024}
\renewcommand{\shorttitle}{Software-based Security Framework for Edge and Mobile IoT}

\begin{document}
\maketitle

\begin{abstract}
With the proliferation of Internet of Things (IoT) devices, ensuring secure communications has become imperative. Due to their low cost and embedded nature, many of these devices operate with computational and energy constraints, neglecting the potential security vulnerabilities that they may bring. This work-in-progress is focused on designing secure communication among remote servers and embedded IoT devices to balance security robustness and energy efficiency. The proposed approach uses lightweight cryptography, optimizing device performance and security without overburdening their limited resources. Our architecture stands out for integrating Edge servers and a central Name Server, allowing secure and decentralized authentication and efficient connection transitions between different Edge servers. This architecture enhances the scalability of the IoT network and reduces the load on each server, distributing the responsibility for authentication and key management.

\end{abstract}

\section{Introduction}
\label{sec:intro}

Given the increasing ubiquity of Internet of Things (IoT) devices in our daily lives, understanding the variety and complexity of cyberattacks these devices face has become imperative. In~\cite{Johnston}, highlights a broad spectrum of vulnerabilities in IoT devices, leading to a large range of attack techniques that compromise data integrity and confidentiality, significantly threatening IoT systems' functionality and security.

One of the primary challenges in the IoT landscape is balancing the affordability of devices with implementation of robust security measures. For instance, jamming and adversarial attacks, which aim to disrupt wireless communication, are particularly disruptive in resource-limited devices~(\cite{dong2013mitigating}). Measures such as signal strength analysis and packet delivery rate verification have been proposed as effective solutions to detect such activities~(\cite{Xu2005}).
Additionally, the insecure setup of IoT devices exposes them to various vulnerabilities. Researchers have suggested introducing artificial noise and robust authentication processes to secure devices during this critical phase~(\cite{Chae2014}). Similarly, for low-level Sybil and spoofing attacks, channel-based detection and RSS consistency emerge as fundamental preventive measures~(\cite{Xiao2009}).

Insecure physical interfaces also represent a considerable risk. Strategies for preventing physical tampering and secure access service management are essential to mitigate these risks~(\cite{Khatri}). For sleep deprivation attacks, which aim to deplete devices' energy, cluster-based techniques and anomaly detection offer a potential solution~(\cite{bhattasali2011survey}).
Fragmentation in IoT devices creates additional vulnerabilities, where replay attacks can be particularly damaging. Content chaining schemes and split buffer approaches are presented as viable solutions~(\cite{Buttyan}).
Within the scope of IoT attacks, there is a need for a deeper analysis of middleware security. Robust authentication, access control, and secure protocols like HTTPS and XMPP are proposed measures to bolster security at this layer~(\cite{Conzon}).

Given the wide range of potential attacks and solutions, it is clear that IoT device security is continuously changing, requiring a comprehensive approach that considers effectiveness, cost, and practicality.

This work-in-progress delves into the intricate challenge of having a balanced approach to security while maintaining operational efficiency and safeguarding against potential threats. 
We propose an architectural framework for secure IoT communications designed to address the emerging challenges of the mobile Internet of Things (MIoT). This architecture is focused on enhancing security through decentralized authentication, which is achieved by integrating a combination of Edge servers and a central Name Server, aiming to provide robust and decentralized authentication mechanisms across the network.

This work extends the software-based security approach proposed in~\cite{soft-based-jferreira} by adding new features for mobile IoT devices and multiple Edge servers, allowing mobile-embedded IoT devices to change their Edge server.
Considering the typical constraints of IoT devices, our architecture optimizes resource and energy consumption using a strategic cache system implemented at each Edge server. This minimizes the need for repetitive and resource-intensive re-authentication processes, thereby conserving energy.
Scalability and flexibility is another feature of the proposed architecture. The design is intended to be highly scalable, easily accommodating an increasing number of IoT devices without compromising performance. Central to this scalability is the architecture's flexibility in seamlessly integrating new Edge servers into the network and allowing devices to switch connections between these servers without friction.
Balancing security with resource usage (efficiency) and energy consumption is vital. The architecture is tailored to balance robust security measures and efficient performance, particularly on resource-constrained IoT devices. 
Given the IoT context of the proposed solution, the encryption mechanism must be lightweight, ensuring minimal resource consumption while preserving data privacy. In alignment with this requirement, the encryption strategy leverages NIST's lightweight encryption standards\footnote{https://www.nist.gov/news-events/news/2023/02/nist-selects-lightweight-cryptography-algorithms-protect-small-devices} to handle data encryption and integrity.

\section{Threat Model and Assumptions}
Developing a threat model and establishing assumptions are crucial to designing a secure architecture. Next, we outline the threat model considered in the design of our architecture, as well as the adopted countermeasures:

\begin{itemize}
    \item \textbf{Unauthorized Access:} Malicious actors are attempting to access IoT devices or communication channels illegally. This threat is addressed by considering robust decentralized authentication through Edge servers and a central Name Server, leveraging secure protocols.

    \item \textbf{Data Tampering:} Attackers try to manipulate or tamper with data during communication. The proposed architecture uses lightweight encryption standards following NIST guidelines for data encryption and integrity.

    \item \textbf{Physical Tampering:} Physical access to IoT devices for tampering or unauthorized manipulation. Secure access service management and regular code integrity verification are applied.

    \item \textbf{Jamming and Adversarial Attacks:} Deliberate disruptions to wireless communication to compromise IoT devices. Signal strength analysis and packet delivery rate verification are used to detect and counter jamming and adversarial attacks.

    \item \textbf{Replay Attacks:} Replay attacks exploiting fragmentation vulnerabilities in IoT devices. Challenge approach, data integrity, and application integrity verifications are used to prevent and detect replay attacks.
\end{itemize}

Considering this threat model, the following assumptions are made:
\begin{itemize}
    \item \textbf{Communication Channels:} The architecture assumes the availability of reliable communication channels, and the threat model focuses on securing the data transmitted over these channels.

    \item \textbf{Adequate Resource Allocation:} The proposed architecture assumes that sufficient resources, such as bandwidth and processing power, are allocated to Edge servers to handle authentication and encryption processes effectively.

    \item \textbf{Cooperation Among Edge Servers:} The architecture assumes a cooperative environment among Edge servers to facilitate seamless authentication and communication between devices switching connections.
\end{itemize}

These threat scenarios and assumptions provide a basis for evaluating the security features of the proposed architecture and guide the implementation of relevant security measures.

\section{Edge and IoT mobile management architecture}

In~\cite{soft-based-jferreira}, we introduced software-driven protection and encryption mechanisms tailored for embedded devices. Our design incorporates an Agent specifically designed to operate seamlessly with low-cost and low-end devices, eliminating the need for any alterations to the underlying hardware. Complementing this, we introduce a Computing Module designed for devices with slightly greater computational capabilities.
The Computing Module empowers devices to write data in secure memory, ensuring ongoing verification of its integrity for sustained protection. Moreover, it leverages the Agents present on the device to strengthen device applications against potential attacks. This is achieved by instructing the Agent to generate a code signature for the application and subsequently validating it, enhancing the overall security posture of the embedded system.

This work extends the previous software-based security approach for mobile IoT devices and multiple Edge servers, allowing mobile-embedded IoT devices to change their Edge server. It aims to balance robust security measures and energy efficiency harmoniously, avoiding new authentication procedures every time a new device changes the Edge server. Our proposed methodology uses lightweight cryptography, strategically optimizing device performance and ensuring security without imposing undue strain on their constrained resources.
The architectural framework incorporates Edge servers and a central Name Server, fostering secure and decentralized authentication processes. This integration ensures seamless and efficient transitions between different Edge servers, enhancing the scalability of the IoT network. Our architecture reduces the burden on individual servers by distributing authentication and key management responsibilities, leading to a more robust and resource-efficient system.

\begin{figure}[t]
    \centering
      \includegraphics[width=0.75\linewidth]{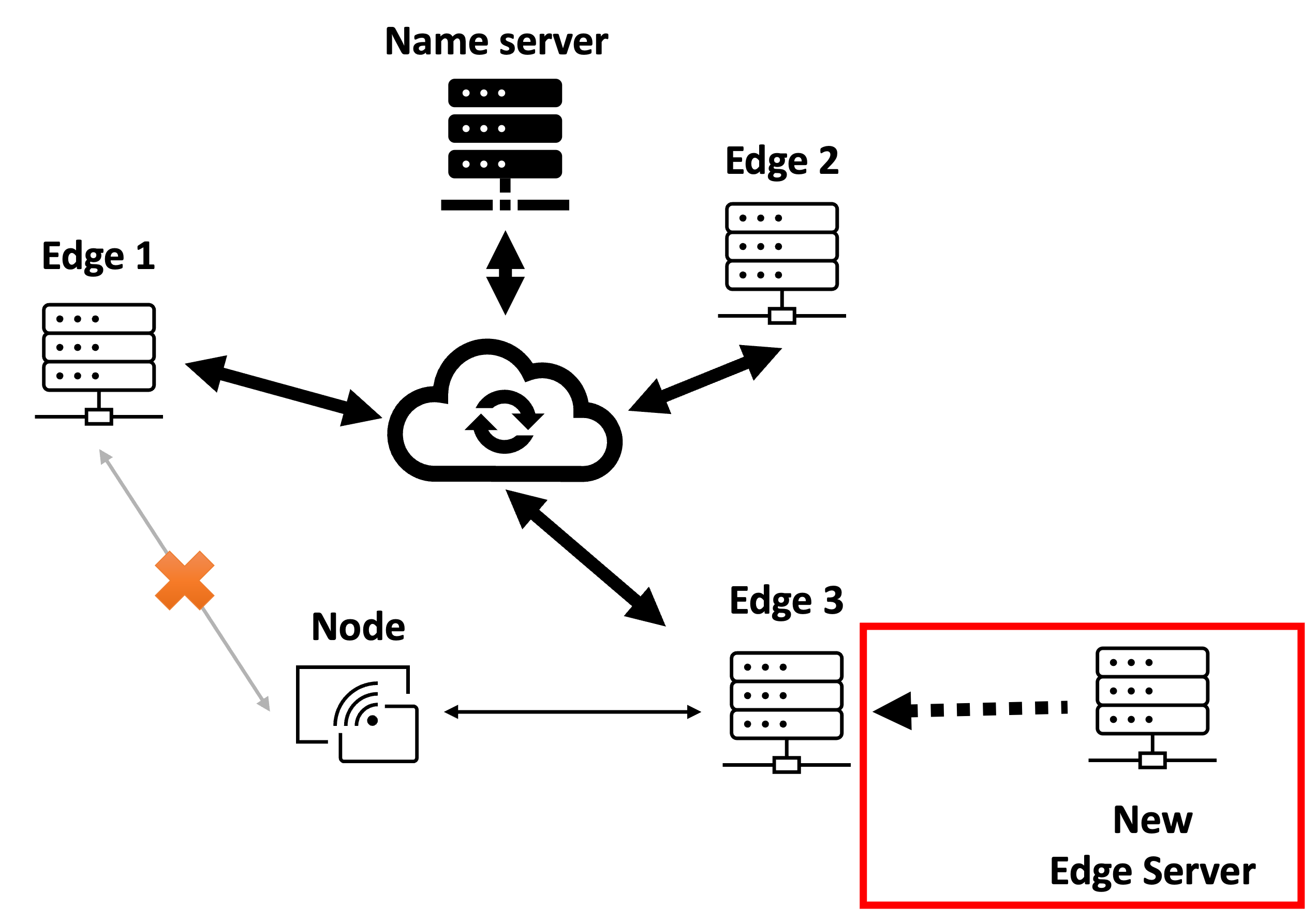}
     \caption{General architecture for Edge and IoT mobile management service.}
     \label{fig:arch}
 \end{figure}

Figure \ref{fig:arch} shows the proposed architecture where we have envisioned a dynamic and mobile network of Nodes capable of disconnecting from one server and seamlessly connecting to another. To facilitate this mobility without needing re-authentication each time a Node switches servers, we propose a system where the authentication credentials, specifically the key and the hashed application data, are transferred alongside the Node to the new server.
To support this seamless transfer, each edge in the network will maintain a secure cache to store authentication information and keys associated with each Node. These keys are subject to renewal, triggered by the Edge servers or when a specific time interval has lapsed, ensuring that authentication remains updated and secure.

Moreover, the key renewal protocol is designed to respect the mobility of the Nodes. If a Node transitions to a new Edge server, the key migrates, and subsequent key renewals are exclusively negotiated with the new server. Upon recognizing that a key is due for renewal and is marked as shared, the initial server will remove that key and associated Node from its registry, effectively transferring full authentication responsibility to the new server.

The system is also designed to achieve high scalability, with provisions for new Edge servers to integrate into the existing system using the protocol proposed in~\cite{10122630}. When a Node wishing to join sends an encrypted authentication message to nearby authenticated Nodes (any Edge server). This message includes a unique identifier and a random key from a pre-established set, ensuring the Node's authenticity from the outset. Upon receipt, an authenticated Node decrypts the message using a default key. If the random key is recognized, a challenge is issued to the new Node to confirm its identity. It uses encryption functions with a mix of the Node's characteristics and randomly generated values. Only upon a successful response to this challenge the process proceed.
This authentication is not isolated. It is part of a broader protocol that seamlessly adds new servers to the network. 

When a new Edge server joins, it undergoes a similar authentication process and must be registered with the name server by the authenticating server. This ensures that all entities within the network, whether Nodes or servers, are authenticated and authorized, maintaining a secure network environment. 

A Server's authentication message includes a random key and the Server's ID, encrypted with a default key known as DK, pre-loaded by an administrator. When an existing authenticated Server receives this message, it decrypts and verifies the random key against its own set. A failure to recognize the key results in the Server's ID being placed on a denylist. However, if the key is validated, the authenticated Server issues a cryptographic challenge to the new Server. 
The challenge and subsequent response ensure that only Servers with the correct credentials can join the network, preventing unauthorized accesses. 

Once authenticated, Nodes can freely move within the network, connecting to different servers without needing to re-authenticate, as their credentials and keys are securely transferred between servers. This procedure aligns with the earlier detailed method for authenticating new servers within the network. It emphasizes the reuse of established authentication, minimizing overhead and enhancing network fluidity. 

When a new Edge server joins and is authenticated, the authenticating Server is responsible for registering the newcomer in the Name Server. 

The Name Server holds essential metadata about the network. This includes mapping addresses to names, the physical locations of Edge servers --potentially using GPS coordinates -- the resources available on each Edge server, and records of which entities have authenticated which.

Each Edge Server and Node implements the software-based security solution outlined in~\cite{soft-based-jferreira}, ensuring that each component contributes to the network's security posture.

When a Node moves from the coverage of one server to another, it relies on the initial server to guide its connection to the optimal Edge server. After consulting the metadata from the name server, the initial server directs the Node to the most suitable server based on a set of criteria derived from the available metadata. This decision-making process is informed by a comprehensive understanding of the network's topology, resources, and authentication relationships, which are meticulously maintained and updated in the Name Server's database.

\section{Device-Edge Communication} \label{sec:Communication}

The proposed architecture defines an innovative protocol for handling Edge servers within the network, allowing it to shift from the traditional central Gateway model to a more flexible and scalable system with multiple Edge servers and a Name Server.

In this system, the authentication of a new Edge server is conducted by an already authenticated Edge server, thus transferring the computational burden and responsibility of the authentication process from a central server to the Edge servers themselves. This decentralized methodology allows each Edge server to authenticate new network participants, simplifying network management and reducing the load on a single central point.

The authentication process begins when an authenticated Edge server sends detailed information about a new Edge server to the Name Server following a successful challenge process. The Name Server then maintains a comprehensive registry of Edge servers in the network, including their addresses, names, locations, resources, and authentication history.

\begin{figure}[t]
    \centering
      \includegraphics[width=0.75\linewidth]{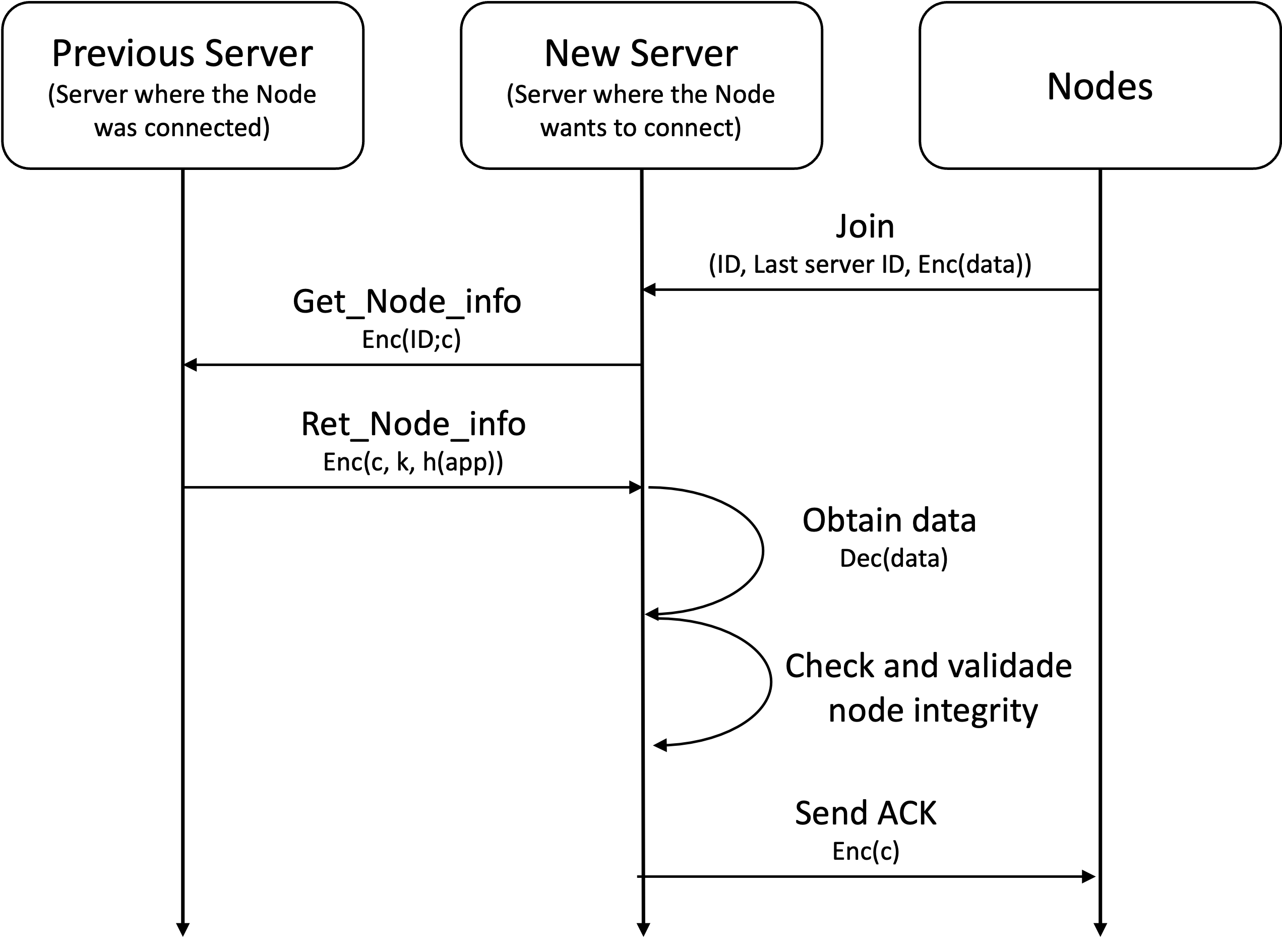}
     \caption{Overview of node connection with a new edge.}
     \label{fig:protocol}
 \end{figure}

Figure~\ref{fig:protocol} illustrates the message exchange process when a node disconnects from a previous Edge server and seeks to connect to another Edge server recommended by the initial one. This connection transition in the proposed architecture is conducted with a particular focus on resource and energy efficiency, thus avoiding unnecessary consumption of both energy and computational resources.

To ensure this efficiency, the previous Edge Server passes the encrypted key to the new Edge server, which was used to establish the initial connection with the Node. It also provides the encrypted hash of the application. As a result, the new Edge Server that will now establish a new connection with the Node does not need to expend energy and computational resources to create a new shared key for authentication.

Upon receiving the key and the application's hash, the new server will decrypt the data sent by the Node and understand its application's hash, which is then compared with the Initial Server's.
If the hashes coincide, it indicates that the data integrity has been maintained, and therefore, the Edge server and the Node can securely establish the connection.

This process not only ensures the security and integrity of the data during connection transitions between Edge servers and nodes but also optimizes the use of resources and energy, a relevant aspect in IoT networks, especially for devices with limited capabilities.

\section{Conclusion}
This work-in-progress is focused on designing secure communication among remote servers and mobile-embedded IoT devices to balance security robustness and energy efficiency. Our focus on addressing several threat vectors led us to propose a comprehensive architectural framework. The integration of Edge servers and a central Name Server, coupled with the deployment of lightweight cryptography, allows for improved security while respecting resource-constrained IoT devices' limitations. 

The strategic use of a secure cache system at each Edge server minimizes the need for resource-intensive re-authentication processes, contributing to energy preservation. The defined threat model allows to establish a robust security framework addressing unauthorized access, data tampering, physical tampering, and other potential threats. We balance security robustness, efficiency, and feasibility by adhering to NIST's lightweight encryption standards and enforcing device participation policies.

This work-in-progress represents a significant advance towards establishing a nuanced and balanced approach to IoT device security. By intertwining decentralized authentication, lightweight cryptography, and thoughtful design principles, our proposed architecture sets the stage for a more secure, scalable, and energy-efficient future for the Mobile Internet of Things (MIoT). 

\section*{Acknowledgments}
This work was supported by the LASIGE Research Unit (ref. UIDB/00408/2020 and ref. UIDP/00408/2020).

\bibliographystyle{unsrtnat}
\bibliography{references}  






\end{document}